\documentclass[aps,twocolumn,prl]{revtex4}
\usepackage{amssymb}
\usepackage{epsfig}
\usepackage{graphicx}
\usepackage{amssymb}
\usepackage{wasysym} 

\begin{document}

\title{Natural versus forced convection in laminar starting plumes}

\author{Michael C. Rogers and Stephen W. Morris}
\address{Department of Physics, University of
Toronto, 60 St. George St., Toronto, Ontario, Canada M5S 1A7 \\
Phone: 416-978-0137, Email: mrogers@physics.utoronto.ca}


\begin{abstract}

A starting plume or jet has a well-defined, evolving head that is driven through the surrounding quiescent fluid by a localized flux of either buoyancy or momentum, or both.  We studied the scaling and morphology of starting plumes produced by a constant flux of buoyant fluid from a small, submerged outlet.  The plumes were laminar and spanned a wide range of plume Richardson numbers Ri.  Ri is the dimensionless ratio of the buoyancy forces to inertial effects, and thus our measurements crossed over the transition between buoyancy-driven plumes and momentum-driven jets. We found that the ascent velocity of the plume, nondimensionalized by Ri, exhibits a power law relationship with Re, the Reynolds number of the injected fluid in the outlet pipe. We also found that as the threshold between buoyancy-driven and momentum-driven flow was crossed, two distinct types of plume head morphologies exist: confined heads, produced in the Ri $ > 1$ regime, and dispersed heads, which are found in the Ri $ < 1$ regime.  Head dispersal is caused by a breakdown of overturning motion in the head, and a local Kelvin-Helmholtz instability on the exterior of the plume.


\end{abstract}


\maketitle

%
%

\section{Introduction}
\label{intro}

Jets and plumes are flow structures of considerable interest~\cite{TURNER,TURjfm} due to their widespread occurrence in industrial and natural systems, from fuel injection~\cite{MATarfm}, to mantle convection~\cite{MORnat, WHIjgr, OLSjfm, GRIpfa}. Both jets and plumes occur when flow discharges from an isolated, submerged source. The distinction between jets and plumes is that a pure jet is driven only by momentum flux at the source, while a pure plume is driven only by buoyancy. If a discharge has a combination of both momentum and buoyancy, there is no sharp distinction between a jet and a plume, and flows span a continuum of possibilities between the two.  Such a flow structure may equally well be referred to as a ``buoyant jet'' or a ``forced plume''; we adopt the latter terminology in this paper.  We experimentally studied the morphology of laminar forced plumes and establish the general scaling of their ascent velocity over the full range spanning pure jets and plumes.  We also identify a sharp transition in the shape of the plume head as the type of forcing is varied.  This transition can be traced to the onset of a divergent flow structure 
in the underbelly of the head. Surprisingly, this transition does not interrupt the scaling of the ascent velocity.

The degree to which buoyancy affects flow emerging from a source is usually quantified by the Richardson number  Ri, defined in Eq.~\ref{Ri_Re_defs}, which is the ratio of buoyancy forces to inertial effects. Thus, Ri can also be viewed as a measure of the extent to which convection is ``forced'', in the Ri~$\ll 1$ limit where inertial effects dominate, as opposed to ``natural'' or ``free'' convection which occurs in the purely buoyant, Ri~$\gg 1$ limit.  In our experiments, slightly buoyant ($< 1\%$ density difference) fluid was injected into an ambient, quiescent fluid of constant density. A wide range of Richardson numbers were accessed by varying the fluids and the flux of buoyant fluid at the source. We measured the ascent velocity and morphology of forced plumes across this range.



Not only do jets and plumes occur in a wide variety of phenomena at various scales, they are also studied in a vast array of geometric configurations and flow scenarios~\cite{GEBarfm, LISTarfm}. We focus here on unconfined flows, not directly interacting with a boundary, and on flows taking place in a quiescent ambient medium of uniform density. By contrast, other important cases are jets and plumes in a crossflow or in a stratified fluid~\cite{LISTarfm}. For a given set of experimental conditions, the most important factor determining jet or plume behaviour and evolution is whether the flow is laminar or turbulent. In addition to being either laminar or turbulent, jets and plumes can be a mixture of both types of flow, with near-field laminar flow that gives way to a far-field turbulent flow~\cite{GEBarfm}. We focused on laminar, forced, compositionally buoyant plumes, for which precise control and characterization of morphology is possible. 

%
%
Forced plumes can be positively buoyant, in the sense that buoyancy forces act in the same direction as the injected momentum, as we consider in this paper, or they can be negatively or neutrally buoyant, as in the pure jet limit.  More complex forced plumes can have several sources of buoyancy, such as in double-diffusive salt fingers~\cite{SCHarfm} or chemically reacting plumes~\cite{ROGprl, ROGpre, CONjfm}. In this paper we only consider buoyancy produced by compositional differences. 
%
%
%
Finally, the state of jet or plume evolution involves one more important distinction: between a steady, well-developed flow and a starting plume, the transient case for which the injected fluid is penetrating the ambient medium and growing in length.  Over the course of its evolution, a laminar starting plume develops its most visually striking feature: a well-defined, evolving head. 

\begin{figure}
\includegraphics[height=12.0cm]{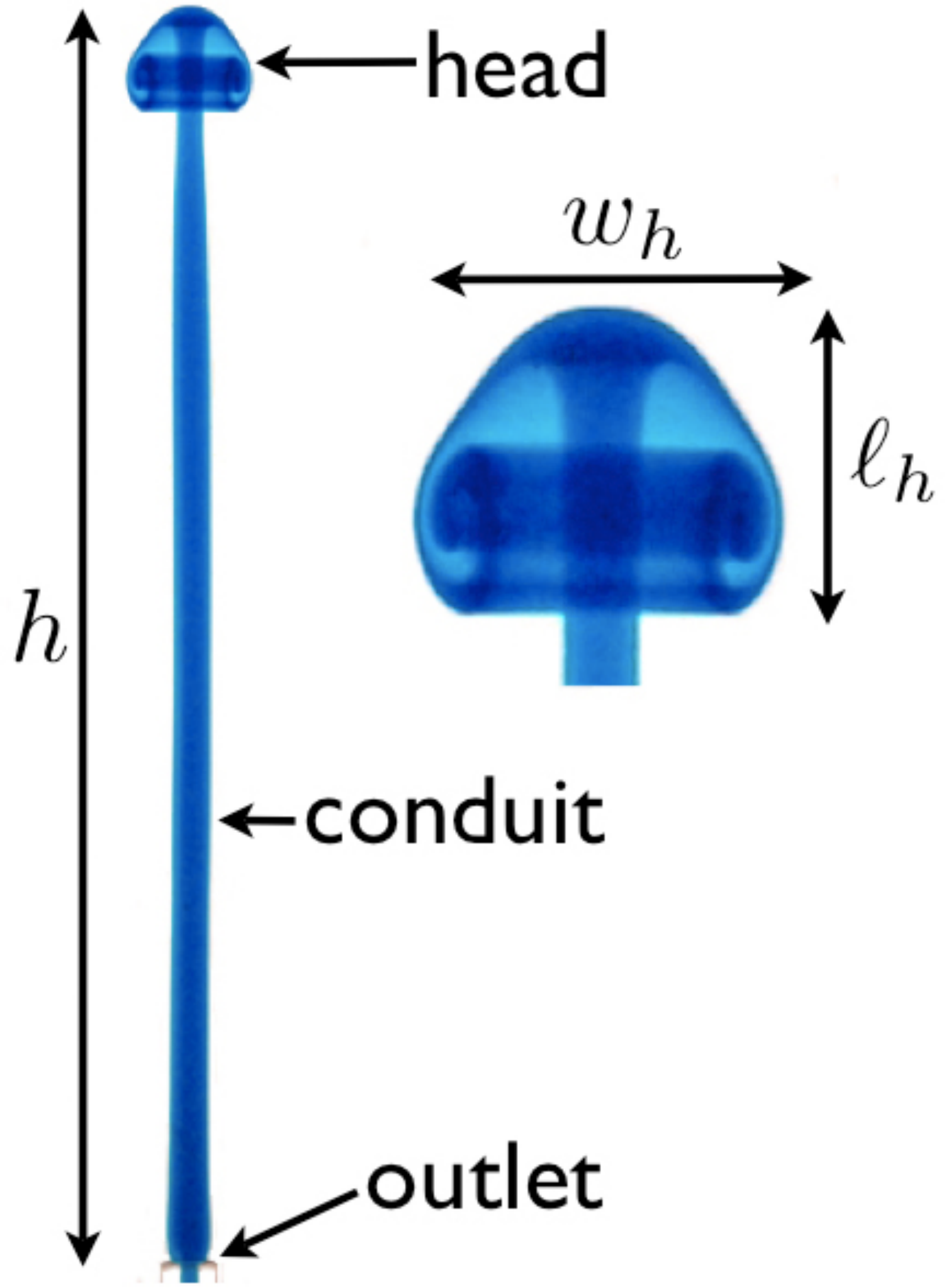}
\caption{\label{plume}
The anatomy of a laminar plume. The entire plume is shown on the left, while a magnified image of the head is shown on the right. $h$ is the height of the plume from the outlet to the top of the head, $w_h$ is the width of the head, and $\ell_h$ is the head length. The plume shown is from set D5 and was created using an injected flow rate of $Q=1.33\times10^{-1}$mL/s. In this image, $h=19.3$~cm, $w_h=1.9$~cm, and $\ell_h=1.6$~cm.}
\end{figure}

Given the wide variety of types of jets and plumes, and the wide array of scientific contexts in which they are studied, it is no surprise that there is a lack of uniformity in the terminology used for jet and plume anatomy. As shown in Fig.~\ref{plume}, we adopt the term {\it head} to describe the structure on top of the plume, and {\it conduit} to describe the fluid emanating from the source that connects and provides volume flux to the head. 
Some authors refer to a starting jet or plume head as a {\it cap}, and the conduit is sometimes called the {\it stem}, or {\it corridor} in the literature. The conduit starts at the {\it outlet} of a small pipe, through which buoyant fluid is being pushed. Finally, we use the term {\it lobe} to describe the bottom, under-turning part of the head.

The most prominent feature of a laminar starting plume is the vortex ring that often forms in the plume head. The generation and evolution of vortex rings has been a subject of longstanding interest in fluid dynamics due to their natural beauty, their utility in engineering applications
and their rich history as simple solutions of the fluid equations~\cite{SHAarfm}. Vortex rings can be produced experimentally by using a cylindrical piston to inject a finite volume of neutrally buoyant fluid into quiescent surroundings~\cite{GHAjfm}. Continuously supplied, neutrally buoyant jets have also been used to study the velocity fields of evolving vortex rings~\cite{SCHmst}. In addition to neutral buoyancy scenarios, vortex ring formation and pinch-off has also been investigated in the context of buoyant starting plumes, where results suggest some universality between free vortex rings produced by pinch-off and vortex rings generated by a piston~\cite{SHUjfm, POTeif}. Buoyant vortex rings may also be created by the pinch-off of accelerating plumes driven by chemical reactions~\cite{ROGprl}.

The majority of work on laminar jets and plumes has focused on the steady variety. Nonetheless, starting jets have attracted interest in studies of combustion~\cite{IGLpof}, and starting plumes in the context of geophysical applications~\cite{MORnat, WHIjgr, OLSjfm, GRIpfa}. Previous work on laminar starting plume heads has focused on ascent velocity~\cite{SHLpof_1, MOSjfm, KAMjfm}, temperature and concentration measurements~\cite{CHApof}, velocity field measurements~\cite{TANpof}, and scaling laws for the head~\cite{MOSjfm}. Since plumes in these studies were driven by a thermal buoyancy flux produced by a localized heater, a relevant dimensionless quantity is the Prandtl number Pr$=\nu/\kappa$, where $\nu$ is the kinematic viscosity and $\kappa$ is the thermal diffusivity. The dependence of plume flow on Pr has been investigated~\cite{KAMjfm}. For plumes driven by compositional differences, the molecular diffusivity $D$ replaces $\kappa$, and the relevant ratio is known as the Schmidt number, Sc$=\nu/D$. Typically, $D \ll \kappa$, so that the diffusion of compositional differences is very slow on the timescale of the evolution of the plume.  This is the case in our experiments, for which $2 \times 10^{3} \leq$~Sc~$\leq 2 \times 10^{6}$.


Diverse plume experiments have yielded a number of plume head morphologies, such as {\it cavity} structures~\cite{OLSjfm, WHIjgr}, {\it umbrellas}~\cite{GRIpfa}, heads that encapsulate a vortex ring, like the one shown in Fig.~\ref{plume}, and heads that do not contain vortex motion in their lobe~\cite{MOSjfm}. Some thermal plume heads become non-axisymmetric under sufficiently high forcing~\cite{SHLpof_2}. More recently, a numerical model of thermally driven 2D {\it line} plumes found four different dynamical regions of starting plume morphology~\cite{MAJpof}.
%


While a variety of head structures are known, the nature of transitions in the type of head that forms as experimental parameters are varied has not been systematically explored. Moreover, the conditions under which a stable laminar head changes its morphology have received little attention.  Similarly, the scaling of the ascent velocity of forced buoyant plumes has not been previously studied.  We address these issues in this paper.

In the next section, we describe the experimental apparatus and protocols, and in the succeeding section, we present the results for the scaling and head morphologies. This is followed by a brief conclusion.



%
%

\section{Experiment}
\label{start_expt}

The apparatus used for our forced plume experiments was a plexiglass tank with a vertical glass capillary tube built into the centre of the tank floor. The inner diameter of the capillary tube, which served as the outlet for fluid injected into the tank, was 3.0~mm, and the inner dimensions of the square tank were 13.4~cm between walls that were 50.2~cm high.  A second, smaller square tank was also used which had 9.8~cm between its inner walls and a height of 33.7 cm. 
A syringe pump was connected to the bottom end of the outlet pipe, allowing fluid to be injected at a controlled, steady rate. The syringe was filled with glycerol-water solutions slightly less dense than the ambient glycerol-water solution into which they were injected. Ambient solutions used in this experiment ranged from a 20\% to an 80\% volumetric ratio of glycerol/water. The densities $\rho_a$ and $\rho_i$ of the ambient and injected solutions were measured using an Anton-Paar densitometer.  Density measurements  were made at the temperature that had been recorded in the room during the respective experiments.  The room temperature was measured to $\pm 0.5^{\circ}~$C, implying a  possible systematic error in the  absolute densities $\rho_a$ and $\rho_i$  of less than $\pm 0.0003$~g~cm$^{-3}$.  However, the difference in densities, $\Delta \rho = \rho_a - \rho_i$, is rather insensitive to the temperature, because of the very small difference in the thermal expansion coefficients of the two fluids.   Since the ambient and injected fluids were nearly identical, the dimensionless ratio of their kinematic viscosities, $\nu_i/\nu_a$, was typically close to unity.  The relevant dimensionless density difference $\Delta \rho/\rho_a$, along with other properties of the experimental fluids, are given in Table~\ref{props}.


%
%
%
%

Two methods of visualization were employed to observe the plumes. In the smaller of the two tanks, shadowgraphy was used. This technique requires a constant light source to be directed horizontally at the tank. By virtue of the difference in the refractive index of the injected and ambient fluids, the less dense (injected) fluid projects a dark image onto translucent white tracing paper attached to the tank at the opposite side to the light source. A CCD camera was used to capture the shadowgram images of the ascending plume. The sets of experiments for which shadowgraphy was used to observe plume behaviour are specified in Table~\ref{props} by set names beginning with ``S". In the larger tank, blue dye was added to the injected fluid for visualization purposes. As fluid was being injected and a growing plume was formed, a digital CCD camera was used to capture images of its evolution. These are given in Table~\ref{props} by set names beginning with ``D". For both the dye and the shadowgraphy experiments, the injected flow rate $Q$ was varied in the range of $3.3 \times10^{-2}$mL/s to $6.67 \times10^{-1}$mL/s. Altogether, 34 forced plumes were analyzed.

%
%
\begin{table*}
\centering   
\begin{tabular}{c c c c c c c}  
\hline\hline                        
Set &  $\rho_i$ & $\rho_a$ & $\Delta\rho/\rho_a \times 10^2$ & $\nu_i \times 10^{-6}$ & $\nu_a \times 10^{-6}$ & $\nu_i/\nu_a$\\ 
 & (g/cm$^3$) & (g/cm$^3$) &  & (m$^2$/s) & (m$^2$/s) & \\ [0.5ex]
\hline                   
S1 & 1.0590 & 1.0599 & 0.0849  & 1.80 & 1.86 & 0.968\\ 
S2 & 1.1151 & 1.1166 & 0.134 & 3.95 & 4.17 & 0.947\\ 
D1 & 1.1745 & 1.1775 & 0.255 & 16.5 & 17.8 & 0.927\\ 
D2 & 1.1752 & 1.1797 &  0.381 & 16.5 & 19.1 & 0.864\\ 
D3 & 1.1897 & 1.1920 & 0.193 & 24.3 & 26.3 & 0.924\\ 
D4 & 1.2045 & 1.2160 & 0.946 & 40.7 & 66.2 & 0.615\\  
D5 & 1.2135 & 1.2161 & 0.214 & 59.9 & 66.2 & 0.905\\   
\hline     
\end{tabular} 
\caption{\label{props}
Fluid properties of the various injected (subscript {\it i}) and ambient (subscript {\it a}) glycerol-water mixtures. 
Densities were measured with a densitometer at the same temperature at which a set of experimental runs were performed. The viscosities 
were interpolated from data given in Ref.~\cite{DOW}.}
\end{table*} 

%
%

\section{Results} 
\label{res}

\subsection{Ascent velocity and scaling }
\label{vel}

\begin{figure}
\includegraphics[height=6.6cm]{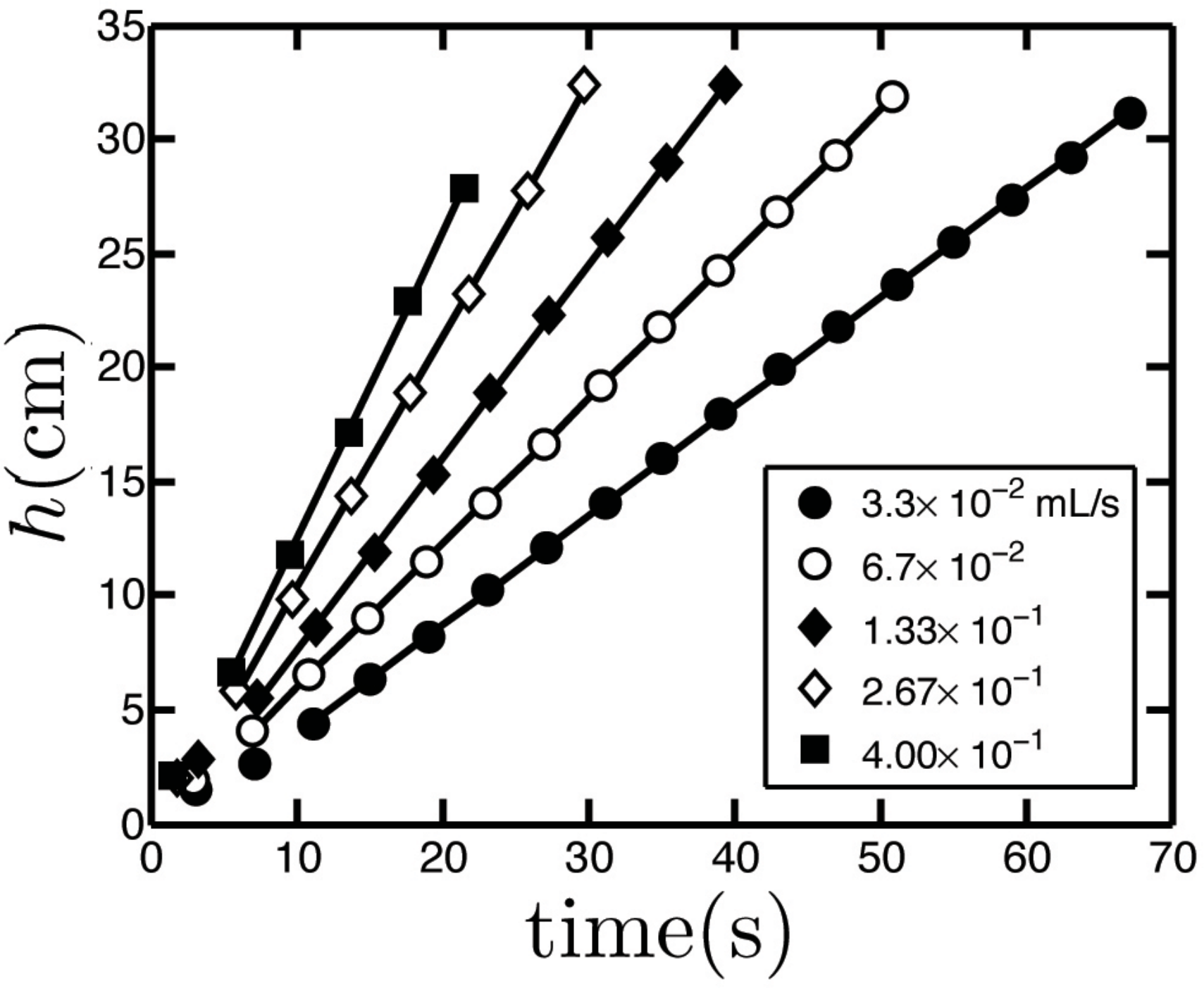}
\caption{\label{height}
The vertical position of the plume head as a function of time for the D4 experimental runs at five injection flow  rates. The solid lines show the linear best fit for plume heights above 4~cm. The slopes of these lines are the constant ascent velocity $v_h$ of each plume.}
\end{figure}

After a short transient, plume heads ascend at a constant velocity. Linear fits of $h$ values extracted from time lapse images were used to determine head velocities, $v_h$, as shown in Fig.~\ref{height}. These fits excluded data in the immediate vicinity of the outlet, which was selected to be $h< 4$~cm. Typically, in the very early stages of plume head formation, the ascent of the head is slower than in the linear regime as plume accelerates towards the constant velocity that it eventually achieves.  
%


As is apparent from Fig.~\ref{height}, for the D4 set of experimental runs, the ascent velocity of the plume head, $v_h$, increases with the injection flow rate $Q$. This trend is consistent for all data sets. To establish the scaling of $v_h$ for all experimental runs, we applied dimensional analysis. We assumed that the flow in the system was dependent on five physical variables: $v_h$, $d$, $\nu_i$, $Q$, and $g'$. Here, $d$ is the diameter of the outlet tube, $\nu_i$ is the kinematic viscosity of the injected fluid, $Q$ is the injected volume flux, and $g'=g\Delta \rho/\rho_a$ is the reduced gravity, where $g$ is the acceleration due to gravity.
Since there are five variables and three fundamental dimensions, a straightforward application of the Buckingham $\Pi$ theorem~\cite{BUCKpr} implies that two dimensionless groups describe the system. Choosing some convenient numerical factors, the dimensionless groups can be written
\begin{equation}
{\rm Ri} =\frac{g'd}{v_h^2}~~~{\rm and}~~~{\rm Re} = \frac{4}{\pi}\frac{Q}{\nu_i d} = \frac{v_{avg} d}{\nu_i}.
\label{Ri_Re_defs}
\end{equation}
Here, Ri is the plume Richardson number, the ratio of the buoyancy forces driving the plume to the inertial terms in the Navier-Stokes equations.  Re is the standard Reynolds number for the flow in the outlet pipe based on its diameter $d$ and average flow velocity ${v_{avg}}$.  We have $Q=Av_{avg}$ in a pipe with the cross sectional area $A= (\pi/4)d^2$.

\begin{figure}[h]
\includegraphics[height=6.6cm]{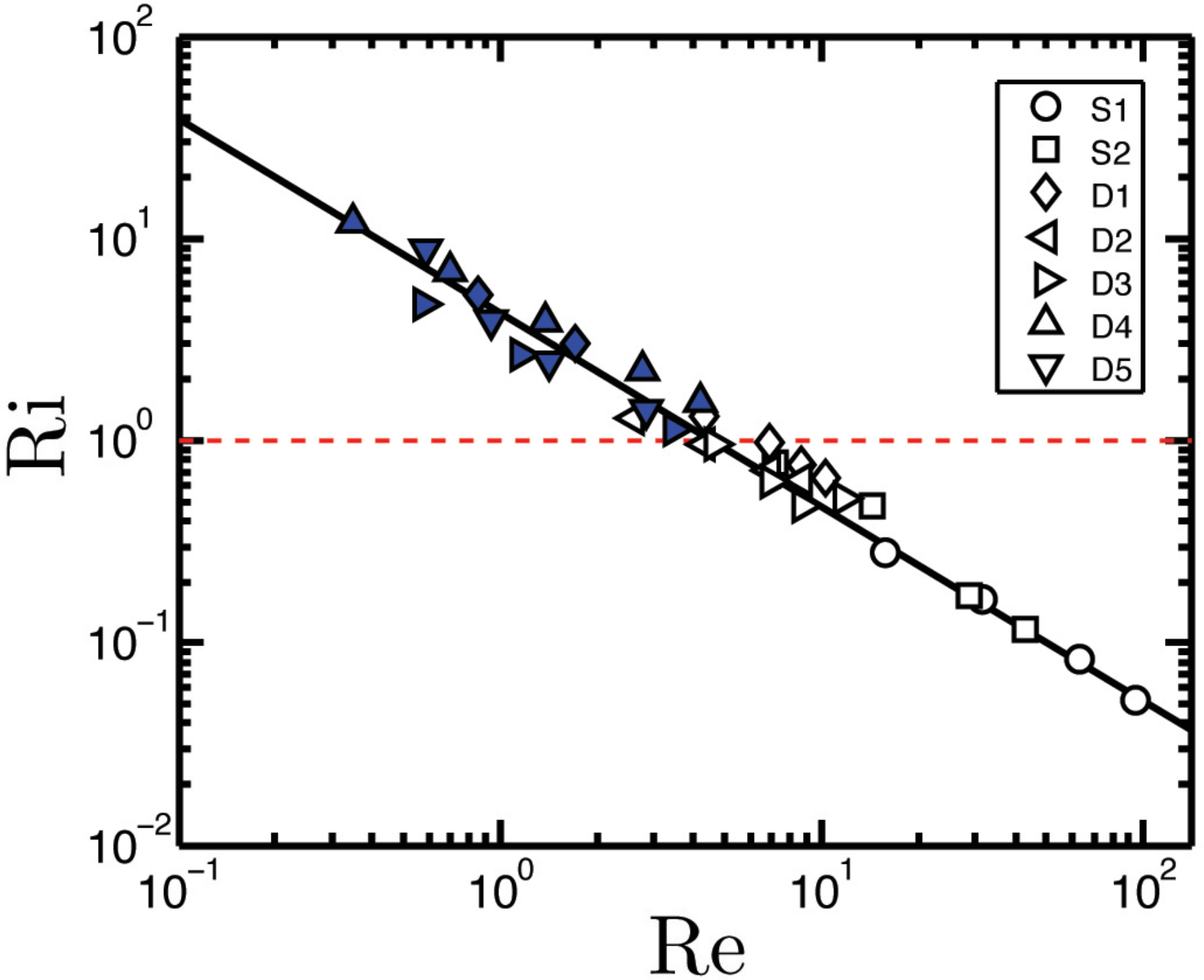}
\caption{\label{rire}
The dependence of the plume Richardson number on the Reynolds number of the injected fluid flow in the outlet pipe. The solid curve is a power law fit to the data, which gives ${\rm Ri}=4.3~{\rm Re}^{-0.96}$. The dashed red line indicates Ri=1. The symbol shading indicates the plume head morphology; shaded symbols indicate confined heads, while open symbols denote dispersed heads.} 
\end{figure}


Plotting Ri vs. Re on logarithmic axes, as shown in Fig.~\ref{rire}, we find that all the experimental data collapse on a single curve. The dependence is well described by a power-law that spans almost three decades in Re more than two decades in Ri.  Re covers a range from well inside the Stokes regime, Re~$\sim 0.1$ to Re~$\sim 100$, but remains small enough that the flow in the outlet pipe is never turbulent. The range of Ri straddles Ri~$\sim 1$, so that we observe both 
flow regimes that are dominated by injected momentum (Ri~$\ll$ 1, for large Re), and plumes in which  buoyancy forces dominate (Ri~$\gg$ 1, for small Re). The power law is of the form
\begin{equation}
{\rm Ri}=a~{\rm Re}^k.
\end{equation}
From a least squares fit of all of the experimental data on Fig.~\ref{rire}, it was determined that $a=4.3 \pm 0.2$ and $k=-0.96 \pm 0.05$.  Thus, we arrive empirically at the remarkably simple result that Ri is approximately proportional to Re$^{-1}$, or equivalently that Ri~Re~= const.  From Eqn.~\ref{Ri_Re_defs}, this implies 
\begin{equation}
 {\rm Ri}~{\rm Re} = \frac{4}{\pi}  \frac{g' Q}{v_h^2 \nu_i } = 4.3 \pm 0.2~,
\end{equation}
{\it i.e.} that all forced compositional plumes are described by a {\it single} dimensionless group that is independent of $d$, the diameter of the outlet pipe.

Isolating $v_h$ in Eqn. 3, we find 
\begin{equation}
v_h =  (0.54 \pm 0.01) \left ( \frac{g' Q}{\nu_i } \right )^{1/2}.
\label{v_h}
\end{equation}
\begin{figure*}
\includegraphics[height=3.2cm]{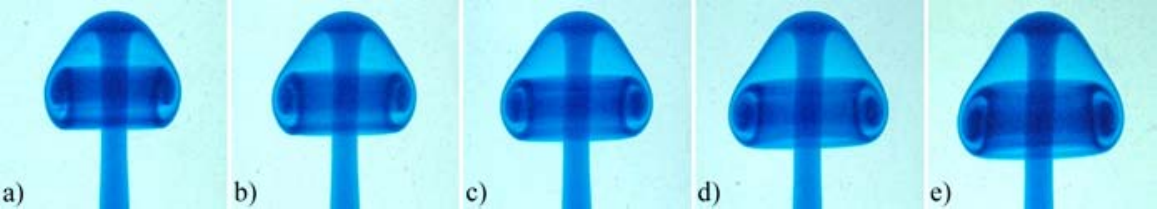}
\caption{\label{plume_sequence80}
The nearly self-similar evolution of a typical confined plume head. Each image is 5s apart. The plume is from set D5 and the injection rate was $Q=2.00 \times10^{-1}$~mL/s.  From the first to the last image in the sequence, $h$ increases from 20.5 cm to 35.2 cm, and the size of the head increases from 1.9cm$\times$ 2.3cm ($\ell_h \times w_h$) to 2.4cm$\times$ 3.1cm.}
\label{ps80}
\end{figure*}
%
%
The above expression for the head ascent velocity $v_h$ may be regarded as a generalization for {\it forced compositional} plumes of the classic scaling proposed by Batchelor~\cite{BATqjrms} for the velocity $v_c$  of purely {\it thermal} plume conduits given by
\begin{equation}
v_c \propto \left ( \frac{g \alpha P}{\nu \rho C_P } \right )^{1/2}~.
\label{v_c}
\end{equation}
Here, the thermal buoyancy flux $g \alpha P/\rho C_P$ replaces the densimetric compositional buoyancy flux $g'Q$ in Eqn.~\ref{v_h}, where $\alpha$ is the thermal expansion coefficient, $P$ is the power input by the heater, $\nu$ is the kinematic viscosity of the isoviscous solution, $\rho$ is the density of the ambient solution, and $C_P$ is the specific heat of the fluid. For the centreline velocity of a steady thermal plume, $v_{cl}$, the proportionality constant in Eqn.~\ref{v_c} is a known function of the Prandtl number~\cite{WORsam}, 
\begin{equation}
v_{cl}=\left [\frac{1}{2\pi}{\log \epsilon^{-2}}\right ]^{1/2} \left ( \frac{g \alpha P}{\nu \rho C_P } \right )^{1/2},
\label{v_cl}
\end{equation}
where $\epsilon$ is a solution of $\epsilon^4 \log \epsilon^{-2}={\rm Pr}^{-1}$.  This result could presumably be generalized to the forced compositional case for which the Schmidt number Sc would replace Pr.

Experiments~\cite{MOSjfm,KAMjfm} on thermal starting plumes in fluids with various Prandtl numbers found that the head rise velocity $v_h$ was related to the conduit centerline velocity given in Eqn.~\ref{v_cl} by
%
\begin{equation}
v_h = (0.57 \pm 0.02) v_{cl}.
\label{ch}
\end{equation}
The prefactor in Eqn.~\ref{ch} is less than one for the obvious reason that the growing plume head must be supplied by a higher speed conduit.  

The physical reason for the scaling of the ascent velocity in each case is clearly the fact that the morphology of the plume eventually becomes independent of conditions near the isolated source of buoyancy flux.  In our case, this implies that $v_h$ should become independent of the diameter of the outlet pipe $d$. Since $d$ is the only length scale in the problem, only a single dimensionless group is required in the overall scaling of $v_h$, as in Eqn.~\ref{v_h}.


%
%
%
%

\begin{figure}
\includegraphics[height=6.6cm]{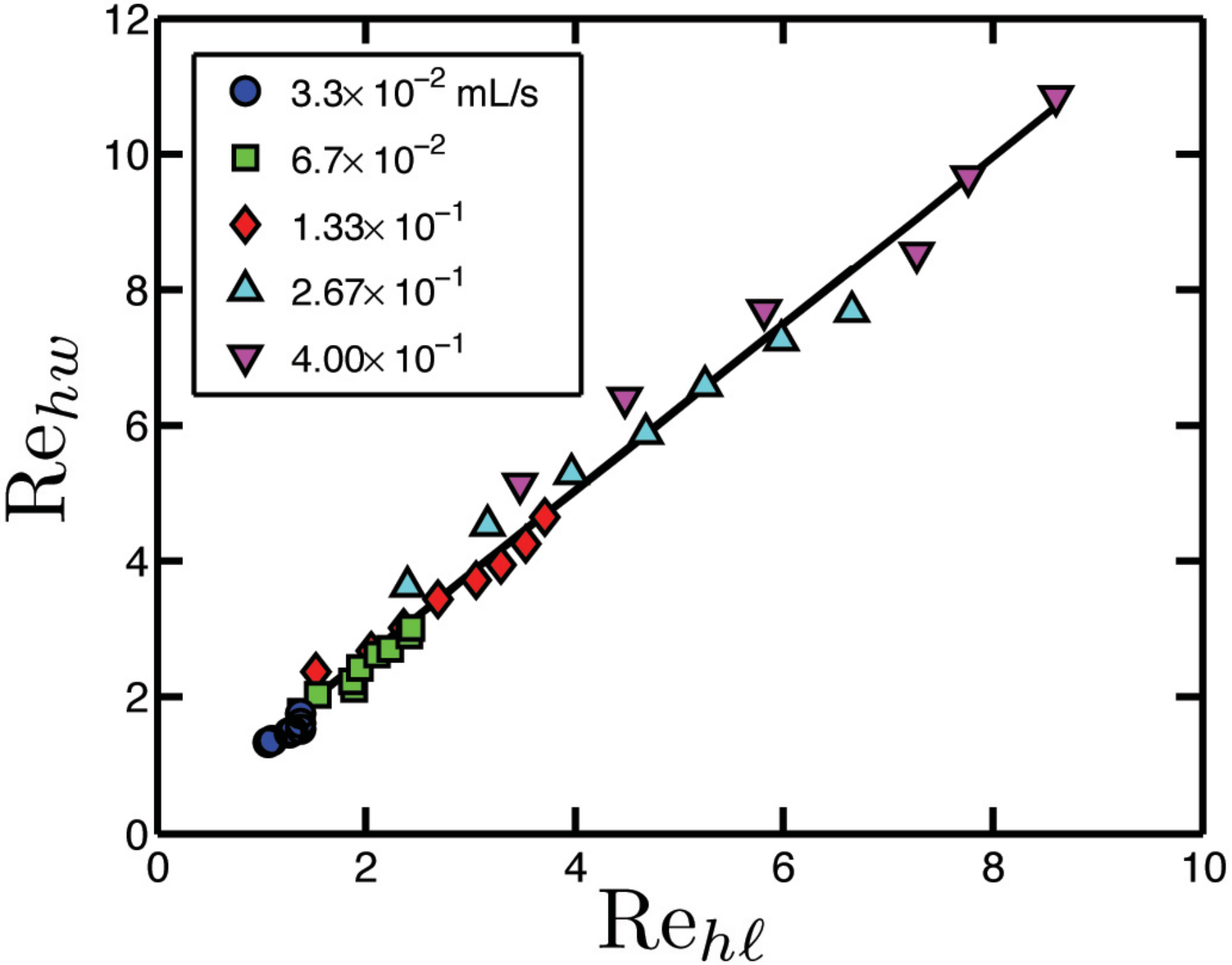}
\caption{\label{D4Re} 
Head width as a function of head length for the D4 set of experiments. All of the plume heads in this set were confined. The head scales $w_h$ and $\ell_h$ have been nondimensionalized as Reynolds numbers ${\rm Re}_{hw}$ and ${\rm Re}_{h\ell}$, respectively. }
\end{figure}


\subsection{Morphology of the plume head }
\label{morph}

The simplicity we found in the previous section for the scaling of the ascent velocity does not extend to the scaling of the plume head morphology.  While the ascent velocity shows the same power-law scaling across a wide range of Re, the head morphology cannot be described by a single characteristic scaling over the same range.
We found instead that there were two distinct types of plume head that exist on either side of ${\rm Ri}\approx 1$.   For lack of an established taxonomy, we classify these head morphologies simply as {\it confined} for Ri~$> 1$ and {\it dispersed} for Ri~$< 1$.    The distinguishing feature is the presence or absence of a stable vortex ring in the lobe of the plume head as it grows during its ascent.  Confined heads, which are observed for larger Ri, exhibit such stable overturning structures, while for unconfined heads, observed for smaller Ri, the vortex ring exists for only a short time before it collapses and disperses.  The domain of each type nicely meet at Ri~$\approx 1$, as indicated in Fig.~\ref{rire}, which forms a reasonably sharp boundary between the two morphologies.   In addition to the stability and size of the vortex ring generated in the lobe, the two types of head have other distinguishing characteristics that are described below. 


\begin{figure*}
\includegraphics[height=9.5cm]{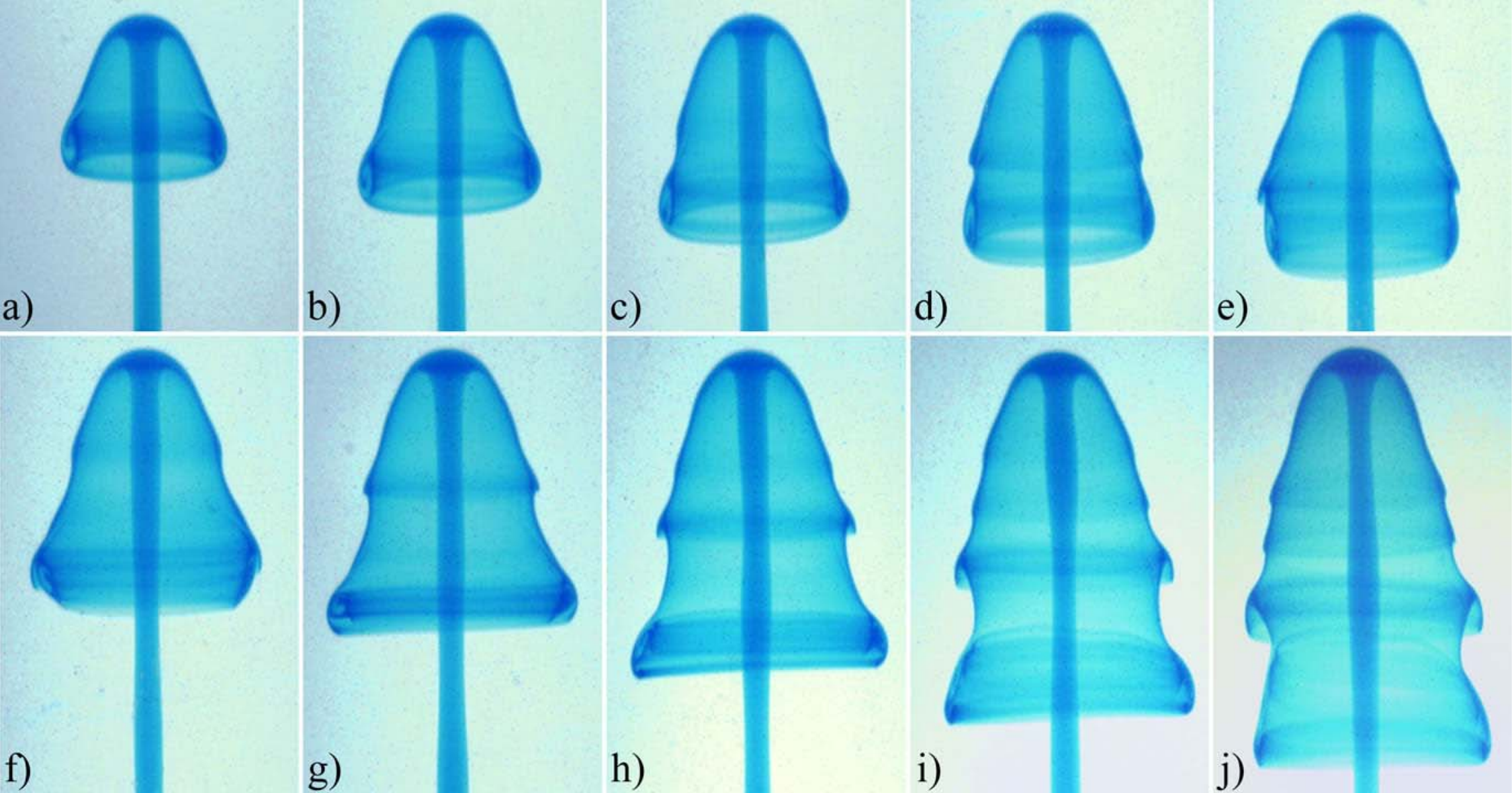}
\caption{\label{plume_sequence65}
A sequence of images of a dispersed plume head during the evolution of a D2 starting plume with $Q=2.67 \times10^{-1}$~mL/s. Each image is 2s apart. From the first
to the last image in the sequence, $h$ increases from 10.4 cm to 31.0 cm. The size of the plume head in a) is $\ell_h =1.6$~cm and $w_h=2.1$~cm. }
\end{figure*}


A confined head is the classic mushroom-shaped laminar plume head. This type of head is shown in Fig.~\ref{plume}, and in a sequence of time-lapsed images 
in Fig.~\ref{plume_sequence80} from a D5 experiment. We use the term ``confined" to describe this type of head because the fluid that comprises it remains within a compact structure. This structure is preserved for the duration of the life of the starting plume, which ends when the head collides with the top of the tank. The fluid in the head
circulates around an axisymmetric vortex ring that remains localized near the top of the plume. As shown in the sequence in Fig.~\ref{plume_sequence80}, in order to accommodate the influx of new fluid delivered to the head through the conduit over time, a confined plume head grows and its vortex ring increases in size. We studied the growth of the head by measuring the width of the head $w_h$, and the length of the head $\ell_h$, as defined in Fig.~\ref{plume}, at various times. These two lengths were nondimensionalized by their associated Reynolds numbers,
\begin{eqnarray}
{\rm Re}_{hw}=\frac{v_h w_h}{\nu_i} ~~~{\rm and}~~~ {\rm Re}_{h\ell}=\frac{v_h \ell_h}{\nu_i}.
\end{eqnarray} 
Figure~\ref{D4Re} shows how the ${\rm Re}_{hw}$ scales with ${\rm Re}_{h\ell}$, for all of the confined D4 plumes. 
We find that ${\rm Re}_{hw}$ is simply proportional to ${\rm Re}_{h\ell}$, independent of $Q$. In all cases, ${\rm Re}_{hw}\sim C~{\rm Re}_{h\ell}$, with $C = 1.24 \pm 0.04$, and hence the aspect ratio of the head is constant, with $w_h/\ell_h = 1.24 \pm 0.04$. This behavior is typical for plume heads in the confined regime. 

\begin{figure}
\includegraphics[height=6.6cm]{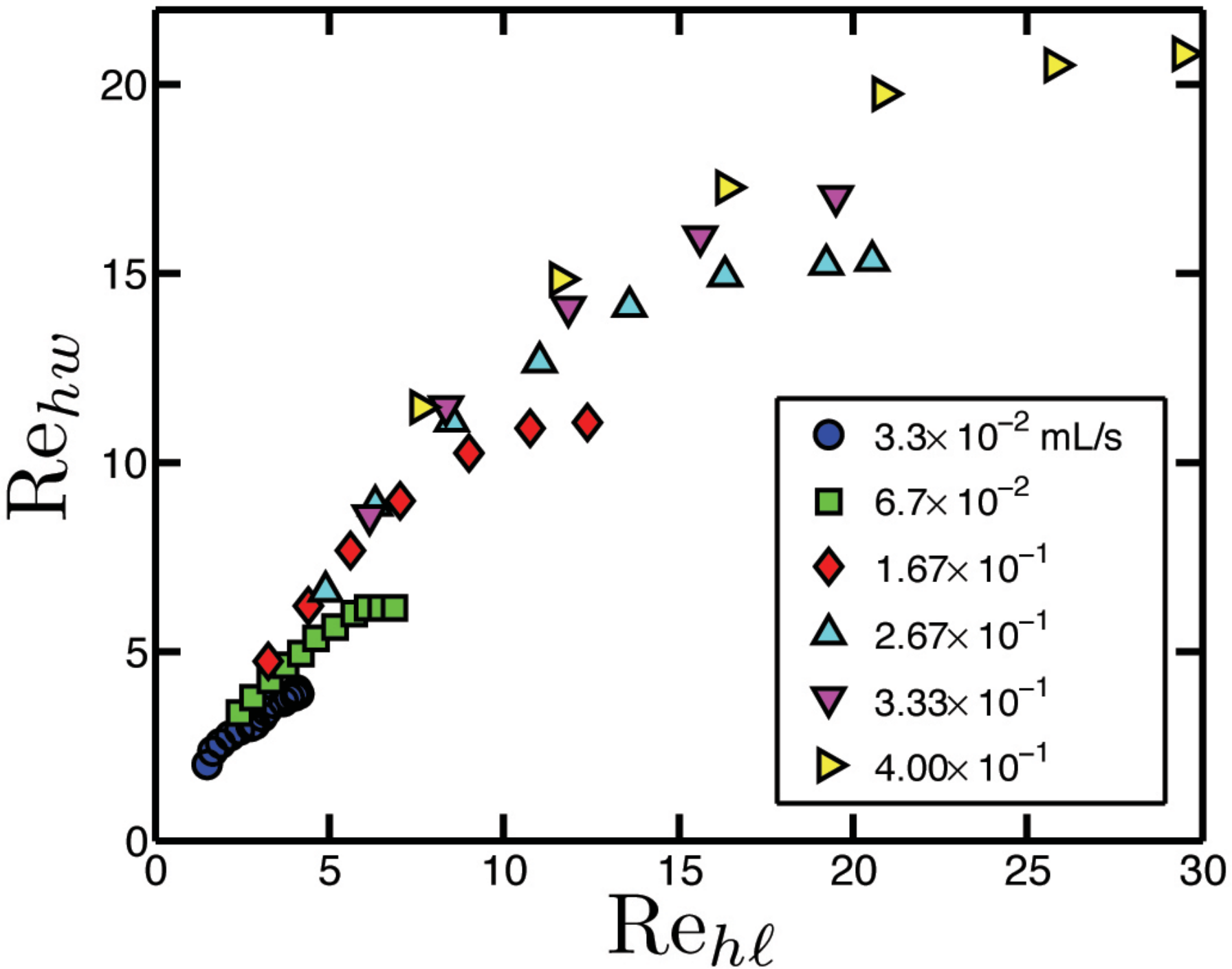}
\caption{\label{D1Re}
Head width as a function of head length for the D1 set of experiments, nondimensionalized as Reynolds numbers. With the exception of the plume with the lowest value of $Q$, all these plumes are dispersed. For all dispersed plumes, the head length grows faster than the width as the lobe becomes unstable. }
\end{figure}

This simple scaling is not found in dispersed plume heads.  The evolution of a dispersed head is shown in Fig.~\ref{plume_sequence65}.  In contrast to confined heads, dispersed heads do not remain compact and do not contain a stable vortex ring structure.  Instead, the height of a dispersed head elongates faster than its width, as shown in Fig.~\ref{D1Re} for D1 plumes.  The head dimensions $w_h$ and $\ell_h$ are only clearly defined in the early stages of growth, after which instabilities in the lobe take over, as shown in Fig.~\ref{plume_sequence65}.

%

\begin{figure}[h]
\includegraphics[height=7.0cm]{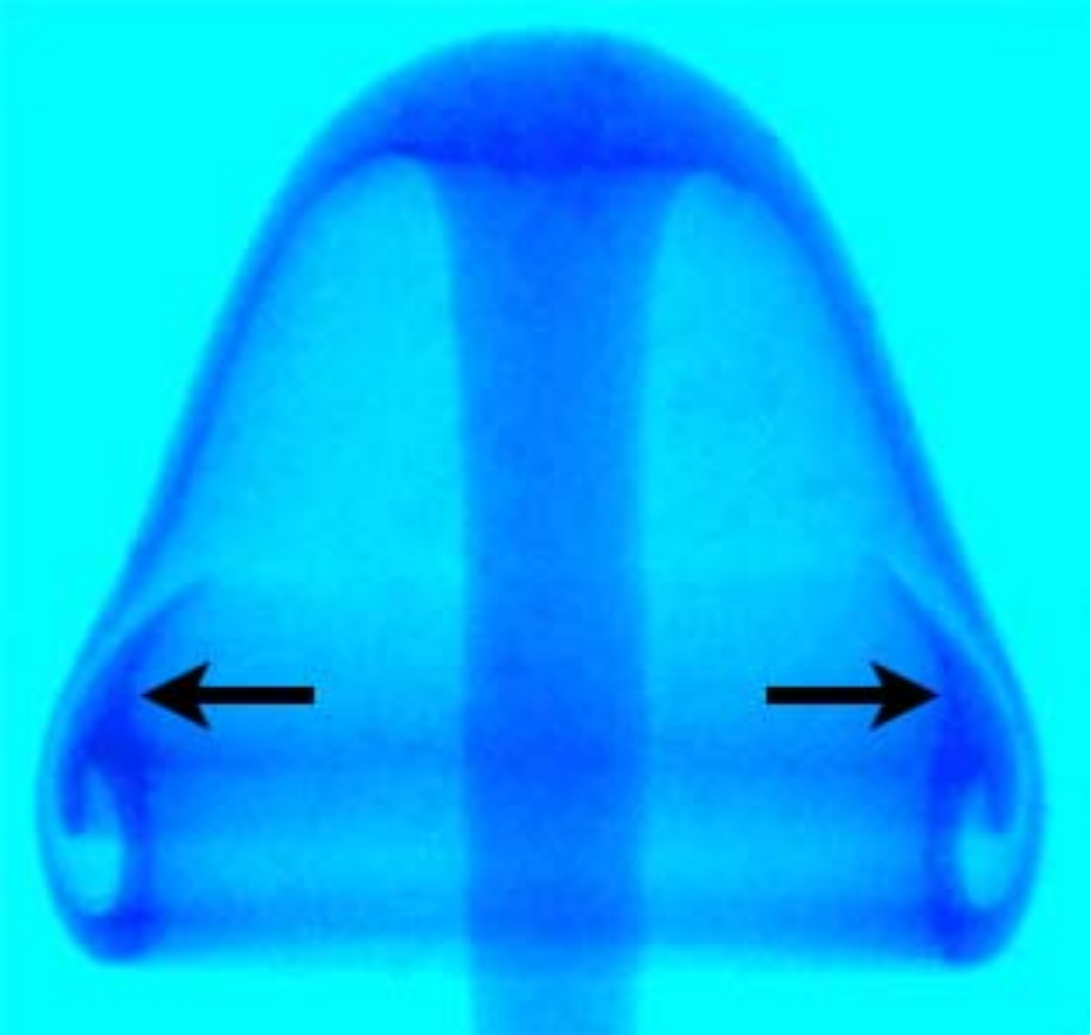}
\caption{\label{preShed}
An image of the axisymmetric hammer-shaped structure (indicated by the arrows) that results from the onset of a divergent flow structure beneath the underbelly of the head.  This structure is not observed in confined heads. The head is from a D3 plume with injection rate $Q=6.67 \times10^{-1}$~mL/s. The dimensions of the plume are $h=21.9$~cm, $w_h=3.7$~cm, and $\ell_h=3.3$~cm.}
\end{figure}

A vortex ring structure forms early in the evolution of a dispersed plume head, but subsequently becomes unstable by an interesting mechanism.  The vortex in the lobe fails to draw in all of the fluid being delivered to it from the outer layer of the head. Instead, the flow diverges and some of the fluid is directed upward and eventually escapes the vortex ring. At the moment when this vortex entrainment breaks down a {\it hammer} shaped structure, unique to dispersed heads, is observed. This structure is barely discernible in Fig.~\ref{plume_sequence65}a,b, and is shown in detail in Fig.~\ref{preShed}. At the site of the hammer structure, the plume head protrudes outwards, forming a bell shape. Above the protrusion, an instability, almost certainly of the Kelvin-Helmholtz type, develops in the outer fluid layers, and the vortex ring becomes cut off from the influx of new fluid. Thereafter, the remnants of the vortex ring near the lobe dissipate, and an elongating fluid skirt that trails below the advancing head is formed. The evolving skirt exhibits further Kelvin-Helmholtz type instabilities, while the overall flow remains axisymmetric, as seen in Fig.~\ref{plume_sequence65}.  A contributing factor to the development these instabilities are small amplitude bulges of fluid that develop in the plume conduit. The bulges develop well above the outlet, and move upward at a higher velocity than $v_h$, causing fluctuations in the volume flux feeding the head. Bulges are visible in the portion of the conduit shown in Fig.~\ref{plume_sequence65} f-j. 

The scenario for dispersed head evolution we have described above is typical of heads that form near ${\rm Ri} \approx 1$. For ${\rm Ri} \ll 1$, the head forms Kelvin-Helmholtz instabilities relatively quickly, before the hammer and bell shapes have time to develop. It should be noted that a Kelvin-Helmholtz instability will develop when the {\it local} shear reaches Ri $ < 1/4$~\cite{TURNER}, where the shear velocity and thickness of the shear layer are the important parameters. It is known from simulations~\cite{MAJpof} that below this critical value of Ri, Kelvin-Helmholtz instabilities appear in a starting plume head. The {\it global} Richardson number we have defined for the whole plume, Eqn.~\ref{Ri_Re_defs}, is not identical to the {\it local} Richardson number that governs the Kelvin-Helmholtz instabilities of the plume head.

\section{Conclusion}

In summary, we have experimentally explored the scaling and morphology of forced compositional plumes, which could  also be called buoyant jets, in the laminar regime.  We focused on starting plumes, for which a well-defined, ascending head exists.  From dimensional analysis, we found that the system is described by the Richardson number of the plume and the Reynolds number of the injected buoyant fluid in the outlet pipe.  We experimentally determined that the Richardson number, which scales with the inverse square of the head ascent velocity, had a simple power law relationship with the Reynolds number, which scales the volumetric flux of the injected fluid. This scaling is exactly such that the diameter of the outlet pipe $d$ drops out.  This reflects the physical fact that the ascending compositional plume's morphology and speed eventually become independent of the details of the localized  source of buoyancy and momentum that produced it.  This result generalizes some previous observations and theory for thermal plumes to the case of forced compositional plumes.  Our results are specific to the case of large Schmidt number and nearly isoviscous plumes. 

The morphology of the advancing plume head exhibits two clear forms, depending on the Richardson number.  For large Richardson number, we observe confined plume heads which contain a stable vortex ring and retain their mushroom shape throughout their evolution.  These plume heads show a simple self-similar scaling.  For small Richardson number, the plume heads become unconfined when the flow feeding the vortex ring fails to close and a thin, trailing skirt is formed.  This skirt undergoes subsequent local Kelvin-Helmholtz instabilities.

It would be interesting to extend this study to the case of forced plumes with both thermal and compositional effects, to non-isoviscous cases and to forced plumes ascending in a density stratified medium.

~~~

\begin{acknowledgments}
MCR thanks Stuart Dalziel and the Department of Applied Mathematics and Theoretical Physics at University of Cambridge. We also thank Mathew Wells, Andrew Belmonte, and L. Mahadevan for useful discussions. This research was supported by the Natural Science and Engineering Research Council (NSERC) of Canada. 
\end{acknowledgments}
  
%
%

\end{document}